# Development and Initial Validation of a Scale to Measure Instructors' Attitudes toward Concept-Based Teaching of Introductory Statistics in the Health and Behavioral Sciences

Hassad, Rossi
Mercy College, New York, USA
**Rhassad@mercy.edu**

Coxon, APM
University of Edinburgh
**Apm.coxon@ed.ac.uk**

**Introduction**

For over a decade, the statistics reform movement has recommended a shift in the teaching of introductory statistics from the predominantly mathematical approach to a more concept-based or constructivist pedagogy (Cobb, 1992; Garfield et al., 2000). Underpinning this reform is a consensus among educators that traditional pedagogy has not been effective in promoting statistical literacy (Cobb, 1992; Moore, 1997; Mills, 2002), a core competency for the evidence-based disciplines such as health and behavioral sciences. In spite of many years of reform efforts, there is growing evidence that students are emerging with a lack of understanding of core concepts, such as statistical variation, chance, graphing of distributions, and sampling distributions (see Delmas et al., 2006). Also, Onwuegbuzi and Leech (2003) observed that many students experience high levels of anxiety about statistics, which can pose "a major threat to the attainment of their degrees".

Such negative outcomes have been attributed to persistence of the traditional pedagogy, that is, "an overemphasis on the mathematical aspects of the subject at the expense of experience with data" (Moore, 2005), fostering rote learning instead of conceptual understanding. Moreover, research from other disciplines has implicated instructors' beliefs and attitudes, as the primary explanation for their failure to use reform-based strategies (Tobin, Tippins & Gallard, 1994; Liljedahl, 2005), however, such research is lacking in statistics education. The only relevant and accessible research report, a case study of statistics instructors, concluded that "the nature and extent of the implementation" of reform recommendations were "often due to the instructor's experience and beliefs about teaching" (Garfileld 2000, p.7). Therefore, the **objective** of this study was to develop and validate a scale to measure instructors' attitude toward concept-based (reform-oriented or constructivist) teaching of introductory statistics in the health and behavioral sciences at the undergraduate level.

**Methodology**

Data were obtained from a maximum variation sample of 227 statistics instructors using an internet-based survey. Established guidelines for scale development were followed (DeVellis, 1991), and structural and psychometric properties were assessed using primarily exploratory factor analysis (EFA), multidimensional scaling (MDS), multiple regression, and hierarchical cluster analyses. Acceptable factorability[1] of the attitude data was established (Bartlett's test of sphericity (p< .001), KMO measure of sampling adequacy = .88). In order to check for criterion validity of the attitude scale, a teaching practice scale was also developed (Table 1). Respondents in the highest quartile (of practice score) were considered high-reform instructors (concept-based or constructivist), and those in the lowest quartile, low-reform instructors (traditional, mathematical or behaviorist).

**Concept-based** (reform-oriented or constructivist) pedagogy was defined as a set of strategies intended to promote statistical literacy by emphasizing concepts and their applications rather than calculations, procedures and formulas. It involves active learning strategies such as projects, group discussions, data collection, hands-on computer data analysis, critiquing of research articles, report writing, oral presentations, and the use of real-life data. **Statistical literacy** is the ability to understand, critically evaluate, and use statistical information and data-based arguments (Garfield, 1999). **Attitude** was conceptualized as an evaluative disposition toward some object, based upon cognitions, affective reactions, and behavioral intentions (Rosenberg & Hovland, 1960; Jaccard & Blanton, 2004).

---

[1] Other assumptions were also met. A pattern coefficient ("loading") of at least .4 was the main criterion for retaining items. (Hair, et al., 1998).

**Results & Discussion**

There were 227 participants (222 reported country), 165 (74%) from the USA, and 57 (26%) primarily from the UK, Netherlands, Canada, and Australia. All taught introductory statistics in health, 94 (41%), behavioral sciences, 102 (45%), or both, 31 (14%) at regionally accredited 4-year institutions in the USA (or the foreign equivalent), and the majority 139 (61%) were male. Age group ranged from 26–30 to 60+ yrs. (median = 41 to 50). Duration of teaching had a mean of 14 years (SD =11, median =10). One hundred and seventy-nine (79%) respondents reported doctoral degrees, and the remainder primarily master's. For degree concentration, 92 (41%) reported statistics, 71(31%) psychology/social/behavioral sciences, 28(12%) health sciences/public health/epidemiology/biostatistics, 19 (8%) education/business, and 17 (8%) mathematics/engineering. Of the 165 instructors from the USA, 135 (82%) identified as Caucasian, and the remainder ethnic minorities.

The attitude scale will be referred to as **FATS** (**F**aculty **A**ttitudes **T**oward **S**tatistics). It consists of 5 subscales (25 items), and explained 51% of the total variance (Table 2), and all (100%) of the common variance. Criterion[2] validity was demonstrated, as high-reform (practice) instructors, on average, reported higher scores (more favorable toward reform) on all attitude measures (Figure 1). All but perceived difficulty were statistically significant ($p < .01$). Teaching practice score was regressed on the subscale scores; and intention ($\beta = .26$, strongest predictor), personal teaching efficacy ($\beta = .24$), and avoidance-approach ($\beta = .20$) emerged as significant predictors ($p < .01$). The model explained 28% of the variance in teaching practice (adjusted $R^2 = .28$, $p < .001$). In another analysis, intention was regressed on the other 4 subscales; and perceived usefulness was the primary predictor of intention ($\beta = .58$, $p < .01$), followed by avoidance-approach ($\beta = .17$, $p < .01$), and personal teaching efficacy ($\beta = .16$, $p < .01$). The model explained 55% of the variance in intention (adjusted $R^2 = .55$, $p < .001$). Perceived difficulty was not statistically significant, and its role in attitude formation, and teaching practice, in this context, is not clear, albeit its moderate relationship with personal teaching efficacy suggests a possible higher-order factor. In general, these findings are empirically and theoretically plausible (see Ajzen, 1991; Venkatesh & Davis, 2000; Armitage & Conner, 2001; Hennessy & Fishbein, 2004; Estrada et al., 2005).

**Table 1: Teaching Practice Subscales (overall Cronbach's alpha = .6)**

| Constructivist (alpha = .66) | *Behaviorist (alpha = .61) |
|---|---|
| I integrate statistics with other subjects. | I emphasize rules and formulas as a basis for subsequent learning. |
| Students use a computer program to explore and analyze data. | I assign homework primarily from the textbook. |
| Critiquing of research articles is a core learning activity. | The mathematical underpinning of each statistical test is emphasized. |
| I use real-life data for class demonstrations and assignments. | I require that students adhere to procedures in the textbook. |
| Assessment includes written reports of data analysis. | I assign drill and practice exercises (mathematical) for each topic. |

Responses (1=never, 2=rarely, 3=sometimes, 4=usually, 5=always). *Items were reverse-coded for the overall teaching practice score, so that higher values reflect more favorable (reform-oriented, concept-based or constructivist) practice. Inter–subscale correlation: Pearson's r = -.06, df=217, ns. An alpha of .6 is a recommended minimum for exploratory studies (Robinson, Shaver & Wrightsman, 1991; Nunally, 1967).

**Table 2: Summary of the Final Factor Analysis Solution**

| Factors/Sub-scales | Observed Inter-Factor correlations (Pearson's r) | | | | | Initial Eigen-Values | % of Total Variance Explained* | Attitude Component | Cronbach's Alpha | N |
|---|---|---|---|---|---|---|---|---|---|---|
| | 1 | 2 | 3 | 4 | 5 | | | | | |
| 1. Perceived Usefulness | | .7 | .49 | .3 | .07 | 8.12 | 16.36 | Cognition | .88 | 227 |
| 2. Intention | | | .49 | .41 | .09 | 2.54 | 13.84 | Intention | .85 | 225 |
| 3. Personal Teaching Efficacy | | | | .39 | .41 | 2.0 | 8.76 | Cognition | .77 | 225 |
| 4. Avoidance-Approach+ | | | | | .11 | 1.34 | 7.08 | Affect | .69 | 225 |
| 5. Perceived Difficulty | | | | | | 1.28 | 5.36 | Cognition | .65 | 226 |
| Overall Attitude Scale (25 items) | | | | | | | 51.00 | Tripartite | .89 | 221 |

Extraction: **Principal Axis Factoring**. Rotation: **Promax.** *The percent of variance explained by each rotated factor was calculated using Cattell's Formula (Cattell, 1978): Variance = structure loading **x** pattern coefficient. +Gray (1970) characterized the mechanism underlying this concept as "facilitative and inhibitory motivational systems", which produce positive and negative **"affect"** respectively. See also, Clark & Watson (1999).

There was no statistically significant difference in attitude with respect to gender, employment status, membership status in professional organizations, ethnicity, highest academic degree, and degree concentration. Instructors (40 years and less) reported significantly lower (less favorable toward reform) levels of overall attitude, perceived usefulness, personal teaching efficacy, and intention, than older instructors (41 –50, and 51 + years). Also, statistically significant (p< .05) but weak relationships were noted between the duration of teaching (years) and personal teaching efficacy (rho = .21), as well as avoidance-approach (rho = .17). This finding suggests a tendency for more experienced instructors to perceive greater capability to use concept-based (or reform-oriented) pedagogy, and exhibit greater approach (less avoidance) in this regard.

---

[2] Other dimensions (content, convergent and discriminant) of construct validity were also established, and details are reported elsewhere.

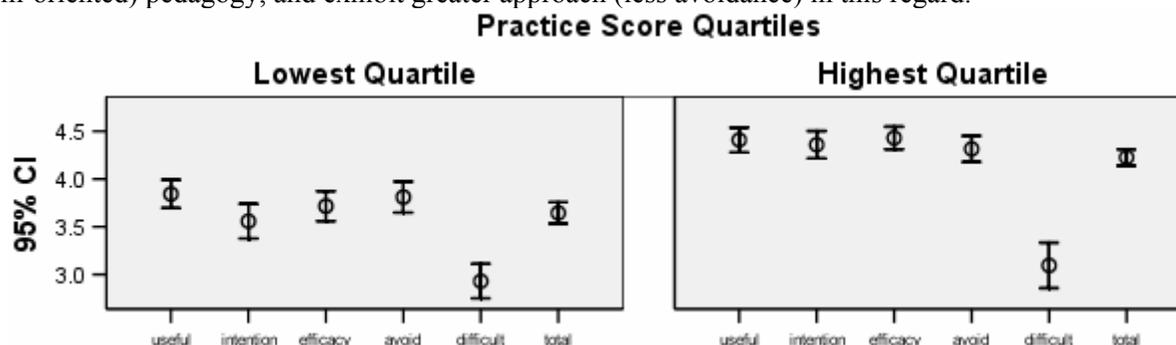

**Figure 1: 95% Confidence Interval of Mean Attitude (Total) and Subscale Scores by Teaching Practice**

**Use of the Attitude Scale**
Subscale scores are calculated based on a 5-point Likert-type scale: 1=strongly disagree, 2=disagree, 3=undecided, 4=agree, 5=strongly agree. Items (with an *) must be reverse-coded so that higher values indicate more positive attitude toward reform-oriented (concept-based or constructivist) pedagogy.

**PERCEIVED USEFULNESS: Beliefs** about the value, benefit or worth of the concept-based approach to teaching introductory statistics. (7 items, alpha = .88)
1. The concept-based approach to teaching introductory statistics (rather than emphasizing calculations and formulas) makes students better prepared for work.
2. The concept-based approach to teaching introductory statistics (rather than emphasizing calculations and formulas) makes students better prepared for further studies.
3. *Emphasizing concepts and applications in the introductory statistics course (rather than calculations and formulas) is a disservice to our students.
4. *The concept-based approach to teaching introductory statistics is for low achievers only.
5. The concept-based approach to teaching introductory statistics enables students to understand research.
6. I am convinced that the concept-based approach to teaching introductory statistics enhances learning.
7. Teaching introductory statistics using the concept-based approach is likely to be a positive experience for me.

**PERSONAL TEACHING EFFICACY: Beliefs** about one's capability to successfully use the concept-based approach to teach introductory statistics. (5 items, alpha = .77)
8. I will adjust easily to teaching introductory statistics using the concept-based approach.
9. *Concept-based teaching of introductory statistics may be problematic for me.
10.*I do not understand how to organize my introductory statistics course to achieve statistical literacy.
11.*Teaching introductory statistics with emphasis on concepts and their applications (rather than calculations and formulas) may be stressful for me.
12. *I am concerned that using the concept-based approach to teach introductory statistics may result in me being poorly evaluated by my students.

**PERCEIVED DIFFICULTY (EASE OF USE): Beliefs** about the effort required to successfully use the concept-based approach to teach introductory statistics. (3 items, alpha = .65)
13.*Teaching introductory statistics with emphasis on concepts and applications rather than calculations and formulas, can be time consuming.
14.*The preparation required to teach introductory statistics using the concept-based approach is burdensome.
15.*Using active learning strategies (such as projects, group discussions, oral and written presentations) in the introductory statistics course can make classroom management difficult.

**AVOIDANCE-APPROACH (Affect):** Positive and negative **feelings**, inclination, proclivity or propensity toward using the concept-based pedagogy to teach introductory statistics. (5 items, alpha = .69)
16. *I am not comfortable using computer applications to teach introductory statistics.
17. Using computers to teach introductory statistics makes learning fun.
18. *I will avoid using computers in my introductory statistics course.
19. I will incorporate active learning strategies (such as projects, hands-on data analysis, critiquing research articles, and report writing) into my introductory statistics course.
20. *I am hesitant to use computers in my introductory statistics class without the help of a teaching assistant.

**BEHAVIORAL INTENTION:** Likelihood of using the concept-based pedagogy to teach introductory statistics. (5 items, alpha = .85)
21. I am engaged in the teaching of introductory statistics using the concept-based approach.
22. I am interested in using the concept-based approach to teach introductory statistics.
23. I want to learn more about the concept-based approach to teaching introductory statistics.
24. *Using the concept-based approach to teach introductory statistics is not a priority for me.
25. I plan on teaching introductory statistics according to the concept-based approach.

**Conclusion**

This multidimensional scale, **FATS** (**F**aculty **A**ttitudes **T**oward **S**tatistics) can be considered a reliable and valid measure of instructors' attitudes toward reform-oriented (concept-based or constructivist) teaching of introductory statistics in the health and behavioral sciences, at the tertiary level. However, additional studies are required in order to be conclusive about this attitude structure and its psychometric properties, and to evaluate for test-retest reliability. It is evident from this research that instructors' beliefs about the usefulness of reform-oriented teaching, as well as personal teaching efficacy are salient in their decision-making process regarding the use of this teaching approach, and these should be the focus of practice change interventions. Consistent with the theory of planned behavior (Ajzen, 1991), perceived behavioral control items (PBC) were included in the instrument, but demonstrated low communality or shared variance, and were removed from the item set. Personal teaching efficacy and avoidance-approach were not initial subscales. This attitude scale (FATS) appears to be the first of its kind. Detailed methodological considerations are presented elsewhere.

**REFERENCES**
**Ajzen**, I. (1991). The theory of planned behavior. Organizational Behavior and Human Decision Processes, 50(2), 179-211.
**Armitage**, C. J. & Conner, M. (2001). Efficacy of the theory of planned behaviour: A meta-analytic review. British Journal of Social Psychology 40, 471-499.
**Cattell**, R. B. (1978). The Scientific Use of Factor Analysis in Behavioral and Life Sciences, New York: Plenum Press.
**Clark**, L. A. & Watson, D. (1999). Temperament: A new paradigm for trait psychology. In L.A. Pervin & O. P. John (Eds.), Handbook of personality: theory and research. 2nd edition. (pp.399-423). New York: Guilford.
**Cobb**, G. W. (1992). "Teaching Statistics," in Heeding the Call for Change, ed. Lynn Steen, MAA Notes No. 22, Washington: Mathematical Association of America, pp. 3-23.
**delMas**, R., Garfield, J., Ooms, A., & Chance, B. (2006). Assessing Students' Conceptual Understanding after a First Course in Statistics. Paper presented at the Annual Meeting of the American Educational Research Association (AERA), April 9, San Francisco, CA.
**DeVellis**, R.F. (1991). Scale development: Theory and applications. Newbury Park: Sage Publications, Inc.
**Estrada**, A., Batanero, C., Fortuny, J. M. & Díaz, C. (2005). A structural study of future teachers' attitudes toward statistics. Paper presented at the Fourth European Conference in Mathematics Education. Sant Feliu de Guissols, Spain.
**Garfield**, J. (1999). Thinking about statistical reasoning, thinking and literacy. Paper presented at First Annual Roundtable on Statistical Thinking, Reasoning and Literacy (STRL-1).
**Garfield**, J. (2000). An Evaluation of the Impact of Statistics Reform. Final Report for NSF project REC-9732404.
**Gray**,J.A.(1970).The Psychophysiological Basis of Introversion-extroversion.Behaviour Research and Therapy, 8,249-266.
**Liljedahl**, P. (2005). Re-educating Preservice Teachers of Mathematics: Attention to the affective domain. Proceedings of the 27th International Conference for Psychology of Mathematics Education - North American Chapter. Roanoke, Virginia.
**Mills**, Jamie D. (2002). Using Computer Simulation Methods to Teach Statistics: A Review of the Literature. Journal of Statistics Education, 10(1).
**Moore**, D. (1997). New pedagogy and new content: The case of statistics. International Statistical Review, 65(2),123-137.
**Moore**, D. (2005). Quality and relevance in the first statistics course. International Statistical Review, 73, 205-206.
**Onwuegbuzie**, A.J., & Leech, N.L. (2003). Assessment in statistics courses: More than a tool for evaluation. Assessment & Evaluation in Higher Education, 28(2), 115-127.
**Robinson**, J. P., Shaver, P. R., & Wrightsman, L. S. (1991). Criteria for scale selection and evaluation. In J. P. Robinson, P. R. Shaver, & L. S. Wrightsman (Eds.). Measures of personality and social psychological attitudes (pp. 1-16). New York: Academic Press.
**Rosenberg**, Milton I., and Carl. I. Hovland (1960). "Cognitive, Affective, and Behavioral Components of Attitudes." Attitude Organization and Change: An Analysis of Consistency among Attitude Components. Ed. Milton J. Rosenberg, et al. New Haven: Yale UP, 1960, 1-14.
**Tobin**, K., Tippins, D. J., & Gallard, A. J. (1994). Research on instructional strategies for science teaching. In D. L. Gabel (Ed.), Handbook of research on science teaching and learning (pp. 45-93). New York: Macmillan Publishing Company.
**Venkatesh**, V., & Davis, F. D. (2000). A theoretical extension of the technology acceptance model: Four longitudinal field studies. Management Science, (46:2), 186-204.

**Acknowledgements: Dr. Edith Neumann and Dr. Frank Gomez**